\documentclass[reprint,amsmath,amssymb,aps,prd,onecolumn,nofootinbib,superscriptaddress]{revtex4-1}
\usepackage{graphicx}
\graphicspath{{fig/}} 
\usepackage[letterpaper,margin=1in]{geometry}
\usepackage{booktabs}
\usepackage{multirow}
\usepackage{pifont}
\usepackage{todonotes}
\usepackage{amssymb}
\usepackage{amsmath}
\usepackage{xcolor, colortbl}
\usepackage{array}
\usepackage[colorlinks = true, bookmarksopenlevel=4,
            linkcolor = blue,
            urlcolor  = blue,
            citecolor = blue,
            anchorcolor = blue]{hyperref}

\begin{document}
\title{The COHERENT Experiment: 2026 Update}
\newcommand{\USDdesc}{\affiliation{Department of Physics, University of South Dakota, Vermillion, SD, 57069, USA}}
\newcommand{\SNUdesc}{\affiliation{Department of Physics and Astronomy, Seoul National University, Seoul, 08826, Korea}}
\newcommand{\FSUdesc}{\affiliation{Department of Physics, Florida State University, Tallahassee, FL, 32306, USA}}
\newcommand{\Dukedesc}{\affiliation{Department of Physics, Duke University, Durham, NC, 27708, USA}}
\newcommand{\TUNLdesc}{\affiliation{Triangle Universities Nuclear Laboratory, Durham, NC, 27708, USA}}
\newcommand{\Mephidesc}{\affiliation{National Research Nuclear University MEPhI (Moscow Engineering Physics Institute), Moscow, 115409, Russian Federation}}
\newcommand{\ITEPnewadesc}{\affiliation{National Research Center  ``Kurchatov Institute'' , Moscow, 123182, Russian Federation }}
\newcommand{\UTKdesc}{\affiliation{Department of Physics and Astronomy, University of Tennessee, Knoxville, TN, 37996, USA}}
\newcommand{\NCSUdesc}{\affiliation{Department of Physics, North Carolina State University, Raleigh, NC, 27695, USA}}
\newcommand{\Sandiadesc}{\affiliation{Sandia National Laboratories, Livermore, CA, 94550, USA}}
\newcommand{\UCASdesc}{\affiliation{University of Chinese Academy of Sciences, Beijing, 100049, China}}
\newcommand{\Tuftsdesc}{\affiliation{Department of Physics and Astronomy, Tufts University, Medford, MA, 02155, USA}}
\newcommand{\ORNLdesc}{\affiliation{Oak Ridge National Laboratory, Oak Ridge, TN, 37831, USA}}
\newcommand{\LANLdesc}{\affiliation{Los Alamos National Laboratory, Los Alamos, NM, 87545, USA}}
\newcommand{\CNLdesc}{\affiliation{Canadian Nuclear Laboratories Ltd, Chalk River, Ontario, K0J 1J0, Canada}}
\newcommand{\IUdesc}{\affiliation{Department of Physics, Indiana University, Bloomington, IN, 47405, USA}}
\newcommand{\ICRRdesc}{\affiliation{University of Tokyo, Institute for Cosmic Ray Research, Kamioka, Gifu, 506-1205, Japan}}
\newcommand{\Okayamadesc}{\affiliation{Department of Physics, Okayama University, Okayama, Okayama, 700-8530, Japan}}
\newcommand{\CMUdesc}{\affiliation{Department of Physics, Carnegie Mellon University, Pittsburgh, PA, 15213, USA}}
\newcommand{\VTdesc}{\affiliation{Center for Neutrino Physics, Virginia Tech, Blacksburg, VA, 24061, USA}}
\newcommand{\Hawaiidesc}{\affiliation{Department of Physics and Astronomy, University of Hawaii, Honolulu, HI, 96822, USA}}
\newcommand{\NCCUdesc}{\affiliation{Department of Mathematics and Physics, North Carolina Central University, Durham, NC, 27707, USA}}
\newcommand{\NCSUnucengdesc}{\affiliation{Department of Nuclear Engineering, North Carolina State University, Raleigh, NC, 27695, USA}}
\newcommand{\WJCdesc}{\affiliation{Washington \& Jefferson College, Washington, PA, 15301, USA}}
\newcommand{\Tokyodesc}{\affiliation{Department of Physics, University of Tokyo, Tokyo, 113-0033, Japan}}
\newcommand{\Kyotodesc}{\affiliation{Department of Physics, Kyoto University, Kyoto, Kyoto, 606-8502, Japan}}
\newcommand{\UFdesc}{\affiliation{Department of Physics, University of Florida, Gainesville, FL, 32611, USA}}
\newcommand{\Concorddesc}{\affiliation{Department of Physical and Environmental Sciences, Concord University, Athens, WV, 24712, USA}}
\newcommand{\SLACdesc}{\affiliation{SLAC National Accelerator Laboratory, Menlo Park, CA, 94025, USA}}
\newcommand{\Laurentiandesc}{\affiliation{Department of Physics, Laurentian University, Sudbury, Ontario, P3E 2C6, Canada}}
\author{M.~Adhikari}\USDdesc
\author{M.~Ahn}\SNUdesc
\author{D.~Amaya Matamoros}\FSUdesc
\author{P.S.~Barbeau}\Dukedesc\TUNLdesc
\author{V.~Belov}\Mephidesc\ITEPnewadesc
\author{I.~Bernardi}\UTKdesc
\author{C.~Bock}\USDdesc
\author{A.~Bolozdynya}\Mephidesc
\author{R.~Bouabid}\Dukedesc\TUNLdesc
\author{J.~Browning}\NCSUdesc
% ...


\begin{abstract}
    The COHERENT experiment measures neutrino-induced recoils from coherent elastic neutrino-nucleus scattering (CEvNS) with multiple nuclear targets at the Spallation Neutron Source (SNS) at the Oak Ridge National Laboratory (ORNL), USA. 
    Several successful CEvNS measurements have been achieved in recent years with tens-of-kg detector masses, with a CsI scintillating crystal, a liquid argon single-phase detector, and high-purity germanium spectrometers. For the next phase, COHERENT aims at high-statistics detection of CEvNS events for precision tests of the standard model of particle physics, and to probe new physics beyond-the-standard model.
    Percent-level precision can be achieved by lowering thresholds, reducing backgrounds, and by scaling up the detector masses. It goes hand in hand with benchmarking the neutrino flux from the SNS. Further detectors will measure CEvNS in additional nuclei, including lighter target nuclei such as sodium and neon, to continue to test the expected neutron-number-squared dependence of the cross section. 
    COHERENT can furthermore study charged-current and neutral-current inelastic neutrino-nucleus cross sections on various nuclei at neutrino energies below $\sim$50\,MeV. Many of these cross sections have never been measured before, but are critical input for the interpretation of core-collapse supernova detection in large-scale neutrino experiments such as DUNE, Super-K, Hyper-K, and HALO.

\end{abstract}

\maketitle






\section{Introduction}

This document was first prepared in response to a request from the Neutrinos \& Cosmic Messengers section of the Update of the European Strategy for Particle Physics and then updated at the beginning of 2026. The COHERENT experiment is sited at the Spallation Neutron Source (SNS) of Oak Ridge National Laboratory (ORNL) in Tennessee, USA. The COHERENT collaboration includes approximately 120 members from institutions in six countries.

\section{Scientific context}
\begin{figure}[htbp]\centering
  \includegraphics[width = 0.48\textwidth]{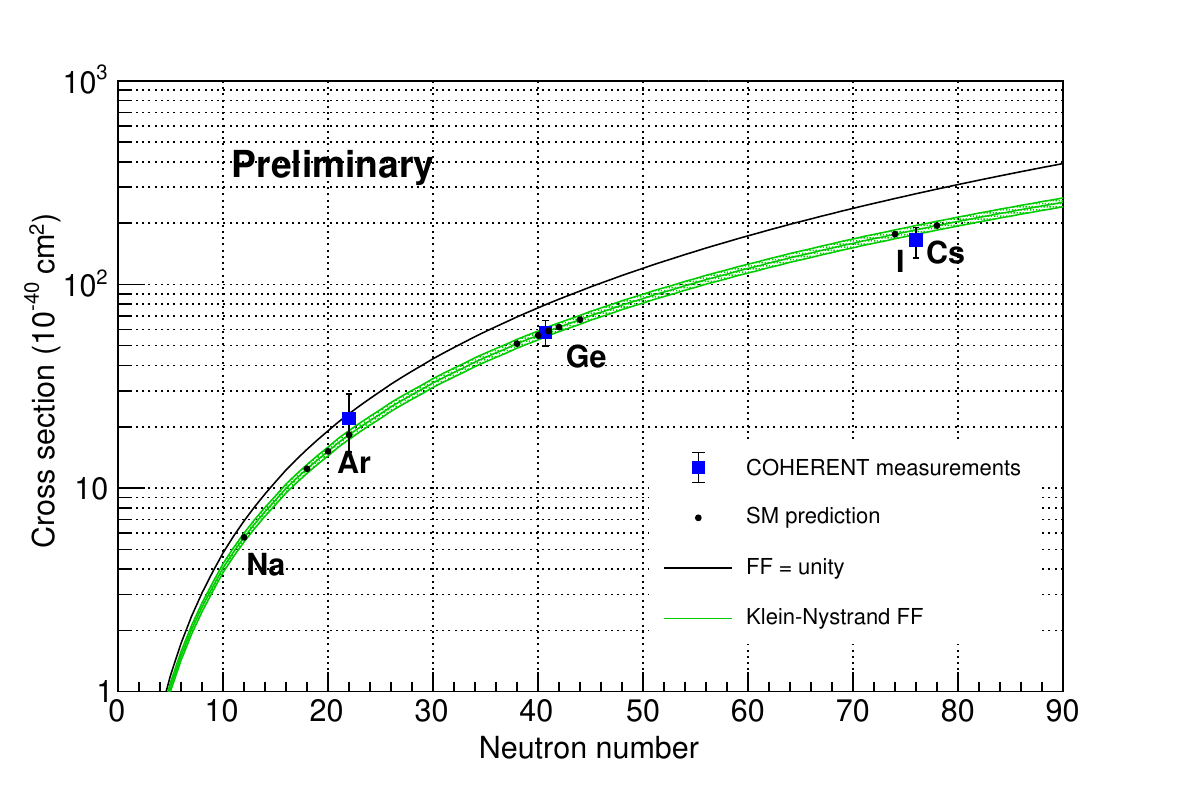}
  \includegraphics[width = 0.48\textwidth]{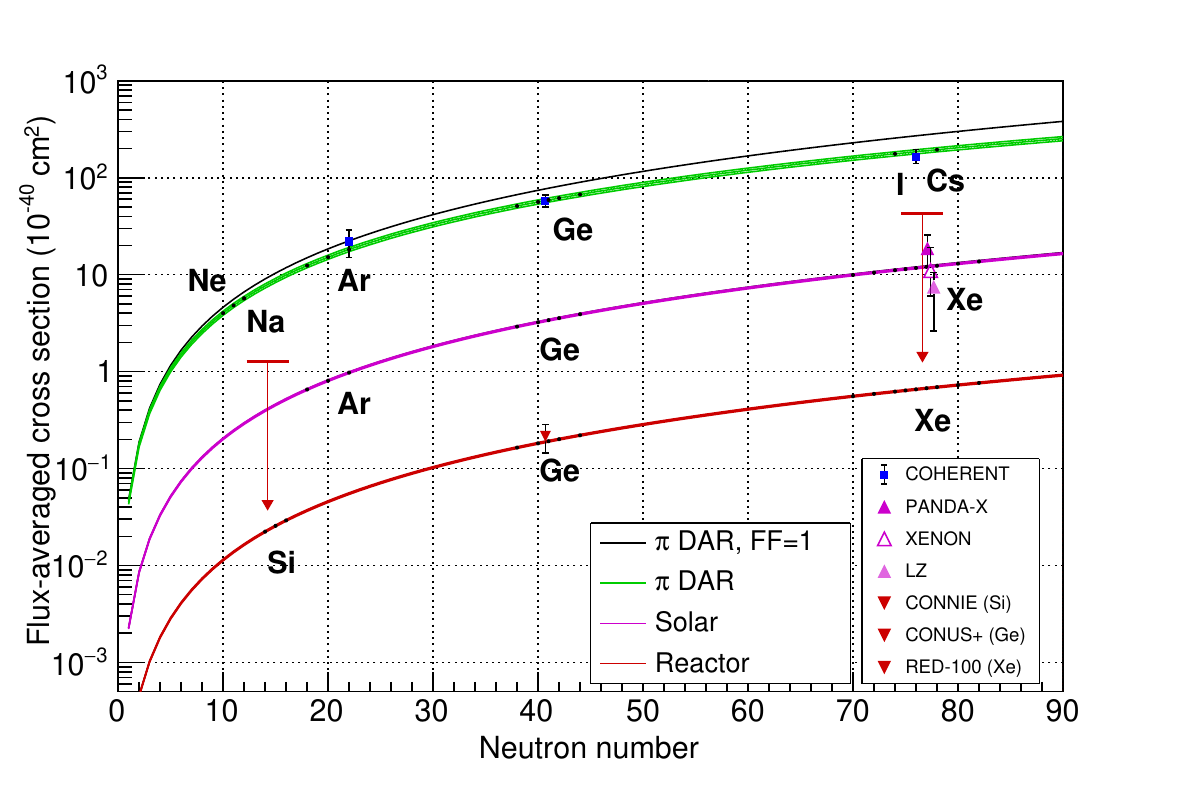}
  \caption{Left: SNS-flux-averaged CEvNS cross section as function of the neutron number of the target nucleus including the impact of the nuclear form factor (FF). The COHERENT CEvNS detections at SNS are marked.  Right: flux-averaged cross section for various neutrino sources (SNS, reactor, solar), including the most precise current detections and upper limits for each neutrino source and target combination.}
  \label{fig:cevnscrosssection}
\end{figure}

Coherent elastic neutrino-nucleus scattering (CEvNS) describes the interaction of the neutrino with the nucleus as a whole~\cite{PhysRevD.9.1389}. The coherence condition holds if the wavelength of the momentum transfer is of the order of or larger than the size of the target nucleus. For medium-sized nuclei, the neutrino energy must be smaller than $\sim$50\,MeV. 
For coherent interactions, the cross section is enhanced by the number of neutrons in the target nucleus squared (see Fig.~\ref{fig:cevnscrosssection});  it is large in comparison to other neutrino interactions such as inverse beta decay~\cite{COHERENT:2017ipa}. 
The high interaction rate and intrinsically small nuclear uncertainties make CEvNS an excellent interaction channel for precision tests of the standard model in modest-scale experiments.

The signature of CEvNS in a detector is the recoil of the scattered nucleus. The detection of these low energy signals results in stringent limits on the noise level of the detector. The maximum recoil energy scales as $\sim \frac{E_\nu^2}{2M}$, where $E_\nu$ is the neutrino energy and $M$ is the mass of the target nucleus.  For neutrinos in the tens-of-MeV range, as available at stopped-pion (pion decay-at-rest) sources, nuclear recoil sensitivity in the few tens-of-keV range is required. Depending on the target material, the recoils result in ionization, scintillation, phonon signals, or a combination of these. It is important to understand the detector energy response;  and dedicated measurements of  the ``quenching" of observable recoil energy are needed to reduce systematic uncertainties (e.g., \cite{COHERENT:2021pcd}, \cite{PhysRevC.110.014613}, \cite{Bonhomme:2022lcz}).

The first detection of CEvNS was achieved at the SNS in 2017 by the COHERENT collaboration with CsI scintillating crystals at room temperature \cite{COHERENT:2017ipa}. COHERENT subsequently detected CEvNS on two more targets with a single-phase liquid argon detector~\cite{COHERENT:2020iec} and high-purity germanium spectrometers (HPGe)~\cite{COHERENT:2025vuz}.   The SNS provides a pulsed beam of stopped-pion neutrinos, as detailed in Section~\ref{sec:nubeam}.

Beyond the efforts at the SNS, CEvNS was very recently also observed at a reactor site.
As power reactors are steady-state neutrino sources, much more effort is required to reduce and understand the background that can only be evaluated during the sparse reactor outages.
Reactor neutrinos have energies below 11\,MeV \cite{an2022first}. The lower energies in comparison to the SNS result in respectively lower recoil energies and consequently detectors with even lower noise threshold by one to two orders of magnitude are required for a successful detection.  The first detection of reactor-neutrino CEvNS at 3.7\,$\sigma$ \cite{Ackermann:2025obx} was achieved by the CONUS+ experiment in 2024 with HPGe detectors. 
Many reactor experiments employ HPGe detectors, which have excellent energy resolution and low thresholds~\cite{Bonet:2020ntx,CONUS:2024lnu,nGeN:2025hsd,TEXONO:2024vfk}. Additional efforts are underway including charged-coupled devices \cite{CONNIE:2024pwt,CONNIE:2024off} and cryogenic calorimeters \cite{NUCLEUS:2021geb,Ricochet:2023nvt}, which are able to detect recoils at even lower energies, as well as cryogenic liquids \cite{RED-100:2024izi} and NaI scintillating crystals \cite{NEON:2022hbk}. Complementary detections of CEvNS at reactors and with stopped-pion neutrinos is highly desirable to study the impact of nuclear effects at different momentum transfers (nuclear form-factor (FF) effects are responsible for the difference between the black and green curves in Fig.~\ref{fig:cevnscrosssection}).

Furthermore, very recently first evidence on CEvNS from solar $^8$B neutrinos was observed by the XENONnT~\cite{XENON:2024ijk}, Panda-X~\cite{PandaX:2024muv} experiments, and now LUX-ZEPLIN~\cite{LZ:2025igz}. These experiments look for recoils of WIMP dark matter with dual-phase liquid xenon time projection chambers and tonne-scale detector masses. The CEvNS interaction of solar neutrinos is an inevitable background in these searches. Excesses have now been seen with significances of 2.7\,$\sigma$, 2.6\,$\sigma$ and 4.5\,$\sigma$ respectively and with several tonne-years of exposure. As dark matter experiments become more sensitive, precise independent CEvNS measurements by COHERENT will enable correct evaluation of this background and potential beyond-the-standard-model signals~\cite{AristizabalSierra:2021kht}. 

Beyond CEvNS, COHERENT at the SNS has excellent sensitivity to charged- and neutral-current inelastic cross sections in the tens-of-MeV range on multiple isotopes. These inelastic interactions have much higher-energy observables, as the neutrino imparts a large fraction of its energy to the final state-- but also have interaction rates per detector mass an order of magnitude or more smaller than CEvNS.   
Few measurements of these cross sections currently exist~\cite{Formaggio:2013kya}.  These interactions are especially relevant in the interpretation of supernova neutrino detection in large-scale neutrino observatories such as the Deep Underground Neutrino Experiment (DUNE)~\cite{DUNE:2023rtr}, Super-Kamiokande~\cite{Super-Kamiokande:2025hxk}, Hyper-Kamiokande~\cite{Hyper-Kamiokande:2021frf}, and HALO~\cite{Zuber:2015ita}. COHERENT has already successfully determined the charged-current cross section on $^{127}$I with a 5.8\,$\sigma$ measurement~\cite{COHERENT:2023ffx} and has measured neutrino-induced neutrons on lead~\cite{COHERENT:2022fic}. Additional measurements on argon (relevant for DUNE), on oxygen (relevant for Super-K and Hyper-K), and on lead (relevant for HALO) are planned. 

\section{The ORNL SNS Neutrino Source}\label{sec:nubeam}

COHERENT takes advantage of the uniquely high-quality neutrino source and well-shielded siting available at the SNS, which produces pulsed $\sim$ GeV protons and collides them on a mercury target.  The primary purpose of this facility is neutron production, but copious neutrinos are a serendipitous by-product. Neutrinos are produced primarily from the decay of pions at rest via the well-understood weak decay sequence that produces monochromatic $\nu_\mu$ on a prompt timescale, followed on a 2.2-$\mu$s muon-decay timescale by $\nu_e$ and $\bar{\nu}_\mu$ with energies up to $\sim$ 50~MeV. Fig. 1 in \cite{Bolozdynya:2012xv} depicts the spectrum divided in neutrino flavor and compared to the neutrino spectrum expected from supernova bursts. The similar energies make the SNS an ideal proxy to study neutrino interactions from supernova neutrinos. The sub-$\mu$s proton pulses result in tight neutrino timing which is highly beneficial for background rejection, translating to an excellent steady-state background suppression factor of 10$^{-3}$ to 10$^{-4}$.  Beam power has been increasing from $\sim$1~MW over the past several years; it is currently in the 1.7~MW range and is expected to reach 2~MW in the near term.  In the further future, a second target station (STS) will be constructed.  One quarter of the total upgraded beam power of 2.4~MW will be directed to the STS; stopped-pion neutrinos from both target stations can be exploited by COHERENT~\cite{sts,Asaadi:2022ojm}.  Expanded facilities at the SNS first target station are also being explored.  Information on COHERENT's detailed flux simulation can be found in Ref.~\cite{COHERENT:2021yvp}.

\begin{figure}[htbp]\centering
\includegraphics[width = 0.8\textwidth]{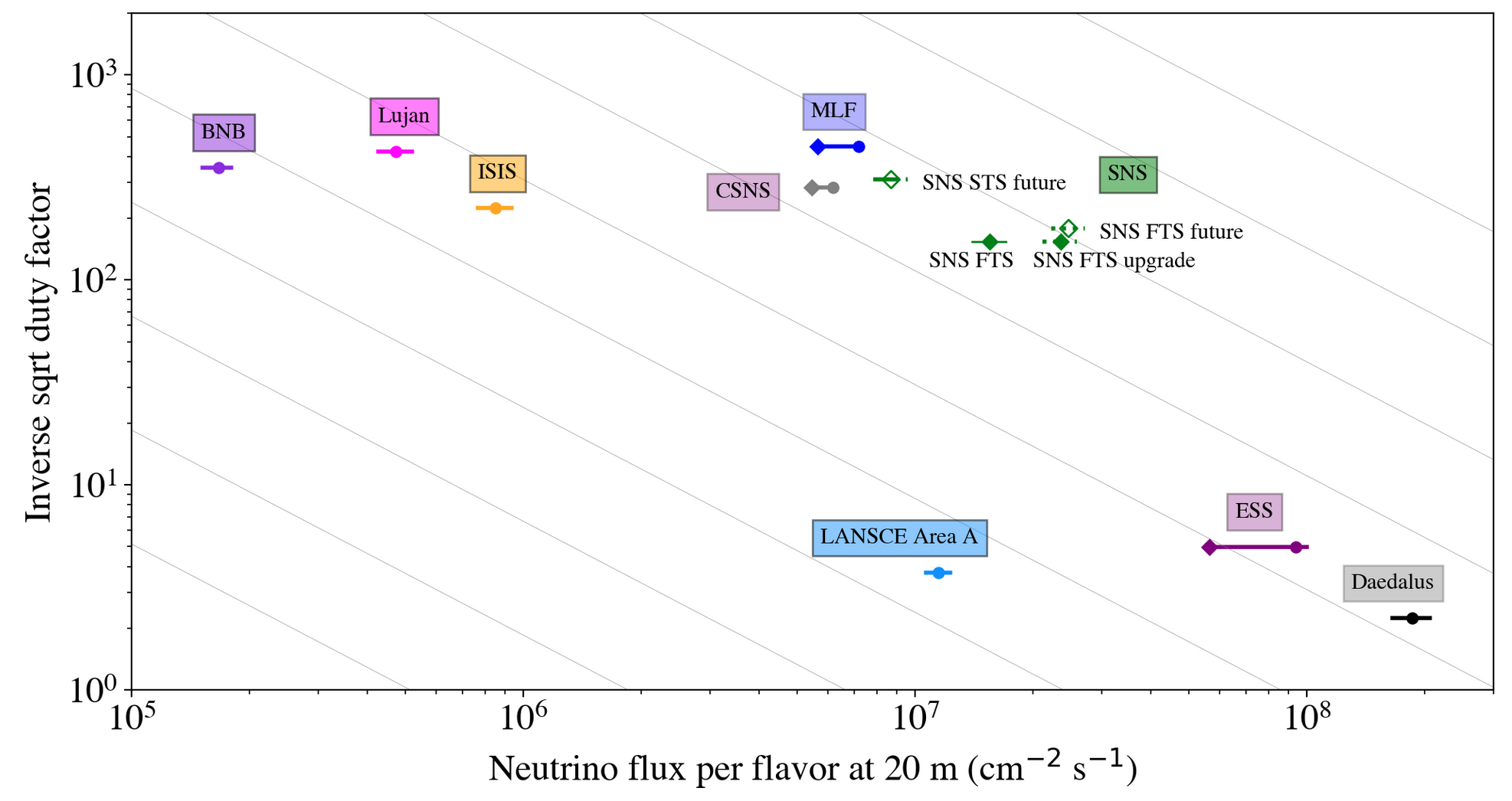}
\caption{Approximate figure-of-merit plot for spallation sources of neutrinos, including past, current and future sources. The $x$ axis is neutrino flux, and the $y$-axis represents the inverse square root of the duty cycle. ``Iso-merit" lines are shown as diagonal lines representing constant signal over square root of steady-state background; further to the upper right means better.  Further details in Ref.~\cite{Asaadi:2022ojm,ks}.}
  \label{fig:spallation-sources}
\end{figure}

Fig.~\ref{fig:spallation-sources} compares spallation sources worldwide for neutrino-physics purposes.  Besides the SNS, some other spallation facilities have existing or proposed CEvNS efforts.  These include: the Lujan Neutron Scattering Center at LANL~\cite{beamLujan,lujanNatlLabSite,CCM:2021leg}, the European Spallation Source (ESS)~\cite{Miyamoto:2023tlr}, the Material and Life science experimental Facility (MLF) at J-PARC~\cite{MLF-beam,Hasegawa:IPAC2018-TUPAL017}, and the Chinese Spallation Neutron Source (CSNS)~\cite{Chen2024}. Future possibilities have also been considered at Fermilab~\cite{Aguilar-Arevalo:2023dai}.

\begin{figure}[htbp]\centering
\includegraphics[width = 0.6\textwidth]{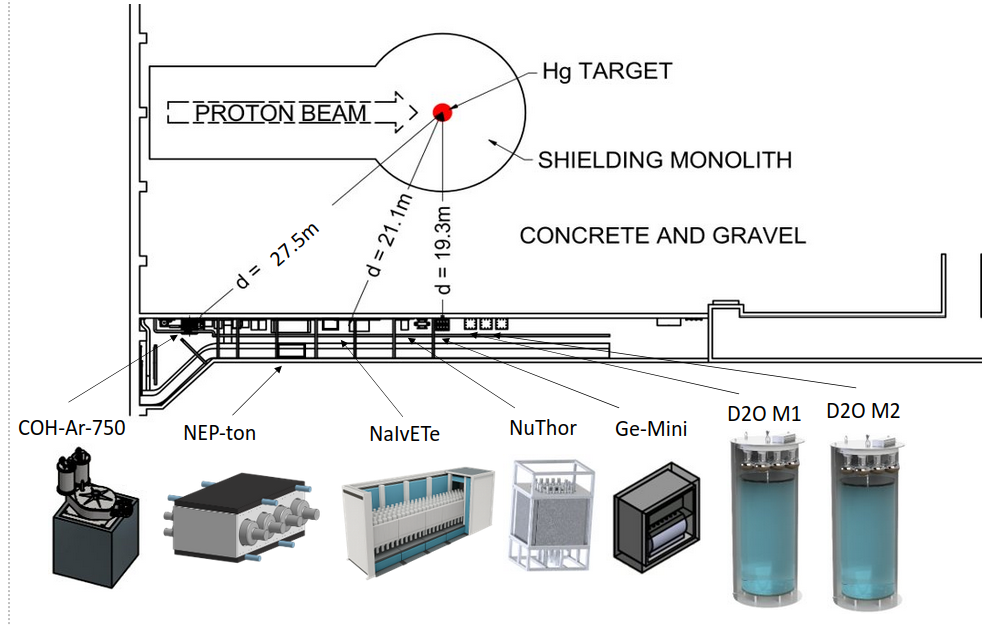}
\caption{Current state of Neutrino Alley in the SNS basement. }
  \label{fig:snsandneutrinoalley}
\end{figure}

\section{COHERENT in Neutrino Alley}
A short overview of COHERENT's subsystems, current and future, is provided in Table~\ref{t:sub}. COHERENT employs multiple detector technologies with different target nuclei spanning a broad range of neutron numbers. 
Beyond CEvNS, neutrino interactions at higher final-state energies can be studied at the SNS. An overview of COHERENT detectors with tens-of-MeV-scale energy sensitivity and sufficient mass to observe reasonable inelastic event rates is provided in Table \ref{t:spd}.
All currently running CEvNS detectors are located in Neutrino Alley in the basement of the SNS with an overburden of 8\,m.w.e. Fig.~\ref{fig:snsandneutrinoalley} depicts the various detectors at their respective distances to the SNS proton target (the source of neutrinos). The basement location is ideal for neutrino detection due to the suppression of the beam-related neutron background by the gravel and concrete between the target and Neutrino Alley. Although this shielding significantly reduces the neutron background, residual beam-correlated neutrons remain a primary background concern for CEvNS measurements (depending on the location within the alley), as they can produce recoil signatures indistinguishable from those of neutrinos. The COHERENT collaboration evaluated this background in detail along the hallway with several neutron measurements with liquid scintillator cells and the mobile MARS detector consisting of scintillator panels interleaved with Gd-painted foil~\cite{COHERENT:2021qbu}. The beam-related background is sufficiently well understood to create little hindrance for CEvNS signal extraction. 

\begin{table}[htbp]\centering\small
\caption{Overview on subsystems for CEvNS detection. \label{t:sub}}
\begin{tabular}{cccccc}
    \hline
    Nuclear & Detector & Target & Distance & Energy threshold& Deployment \\
    target & Technology & Mass (kg) & from source & (keV$^\dagger$)& dates \\\hline\rowcolor{lightgray}
    CsI[Na] & Scintillating crystal & 14 & 20 m & 5 & 2015-2019\\\rowcolor{lightgray}
    Ar & Single-phase LAr$^\star$& 24 & 27.5 m & 20 & 2016-2021\\
    Ge & HPGe PPC$^\ddag$ & 18 & 19.2 m & 2.5 & 2022 - present\\\rowcolor{pink}
    NaI[Tl] & Scintillating Crystal & 3500 & 22 m & 13 & 2022 - present \\\rowcolor{pink}
    Ar & Single-phase LAr$^\star$ & 476 & 27.5 m & 20 & from 2026\\\rowcolor{lime}
    CsI & CsI+SiPM arrays at 77 K & $\sim$6.4  & 20 m & $\sim$0.5 & from 2026\\\rowcolor{lime}
    CsI & CsI+SiPM arrays at 40 K & $\sim$10  & 20 m & $\sim$0.5 & from 2027\\\rowcolor{lime}
    Neon & Single-phase neon &   $\sim$20    &   $\sim$15 m     &  $<$25     & from 2026   \\\hline
\end{tabular}
\\\noindent \colorbox{lightgray}{Completed} \colorbox{pink}{Under deployment} Current \colorbox{lime}{Planned}\\ $^\star$liquid argon, fiducial volume stated, $^\ddag p$-type point-contact, $^\dagger$nuclear recoil energy, approximate threshold
\end{table}

\begin{table}[htbp]\centering\small
\caption{Additional detectors that broaden the physics reach of COHERENT.  \label{t:spd}}
\begin{tabular}{cccc}
    \hline
    Name & Detector Technology & Main purpose & Deployment dates \\\hline\rowcolor{lightgray}
     &  & Measure $\nu_e+$ I CC cross section &  \\\rowcolor{lightgray}
    \multirow{-2}{*}{NaIvE} &  \multirow{-2}{*}{185~kg NaI[Tl] crystals} & \& beam-related backgrounds & \multirow{-2}{*}{2016 - 2024} \\\hline\rowcolor{lightgray}
     NIN  & Liquid scintillator cells & Measure neutrino-induced &  \\\rowcolor{lightgray}
    cubes& in lead and iron shields  & neutrons (NIN) in lead \& iron & \multirow{-2}{*}{2015 - 2022}\\\hline 
    \multirow{2}{*}{MARS} & Scintillation panels inter- & Measure beam-related & 2017 -\\
    & leaved with Gd-painted foils & neutrons in Neutrino Alley & present\\\hline 
    \multirow{2}{*}{NuThor} & Th enclosed by NaI crystals & Measure neutrino-induced& 2022 -\\
    & and Gd-doped water brick& fission on thorium & present\\\hline 
     & heavy water & Measure neutrino flux precisely & \\ 
    \multirow{-2}{*}{D$_{2}$O M1} & Cherenkov detector & \& $\nu_e+$O inelastic cross section & \multirow{-2}{*}{2023 - present}\\\hline \rowcolor{pink}
     & Light ($\rightarrow$ heavy) water & Measure  & \\ \rowcolor{pink}
     
    \multirow{-2}{*}{D$_{2}$O M2 (H$_{2}$O)} & Cherenkov detector &  $\nu_e+$O inelastic cross section & \multirow{-2}{*}{from 2026}\\ \hline\rowcolor{pink}
    NEP-ton  & Cherenkov radiation  & Determination of $\nu_e+$Pb and $\nu_e+$O  & \\ \rowcolor{pink}
     (Pb glass) & in Pb glass blocks &  inelastic cross section & \multirow{-2}{*}{2025-present}\\\hline\rowcolor{lime}
    Pb  & Pb sheets between & Directional $\nu_e$ CC on Pb detection,   & \\ \rowcolor{lime}
    sandwich & scintillator panels  &  forbidden transitions & \multirow{-2}{*}{from 2026}\\
    \hline\rowcolor{lime}
    H$_2$O & Light water & Dedicated $\nu_e+$O & \\ \rowcolor{lime}
     & Cherenkov detector &  inelastic cross sections & \multirow{-2}{*}{from 2026}\\\hline
     \rowcolor{lime}
    LArTPC & Liquid argon time- & Measure $\nu_e+$Ar inelastic & \\ \rowcolor{lime}
     & projection chamber &  cross section & \multirow{-2}{*}{from 2026}\\\hline

\end{tabular}
\\ \colorbox{lightgray}{Completed} \colorbox{pink}{Under deployment} Current \colorbox{lime}{Planned}

\end{table}

\begin{itemize}
\item \textbf{CsI and CryoCsI:} 
The first detection of CEvNS was achieved with a 14.6\,kg doped CsI[Na] crystal at room temperature and an energy threshold of 7.5\,keV$_{nr}$ ($nr$ = nuclear recoil) at a distance of 19.3\,m to the neutrino source.
In total, an exposure of 13.99\,GWh was collected within four years and an overall significance of 11.6\,$\sigma$ was observed \cite{COHERENT:2021xmm}. The result is in agreement with the standard model. The detector observed scintillation light with a photomultiplier tube (PMT). 
The recoil detector response was precisely determined by the collaboration via dedicated quenching factor measurements \cite{COHERENT:2021pcd}, which strongly reduced the systematic uncertainty between the first and second results. 
The first CsI detector has now been decommissioned. To study CEvNS on CsI with an even higher precision, it is possible to leverage the improved light yield of undoped CsI at cryogenic temperatures read out with SiPMs \cite{COHERENT:2023sol}.  
A 0.5\,keV$_{nr}$ threshold at 40\,K is in reach based on COHERENT test-stand studies.
The first phase of this detector aims at the deployment in Neutrino Alley of a detector mass of 10\,kg. 
The lower threshold--the lowest one within COHERENT-- will result in $\sim$10$^3$ CEvNS counts per year of operation and open up opportunities to search for beyond-the-standard-model physics and dark matter at these low energies. 
A prototype cryogenic CsI system operating at liquid nitrogen temperature with traditional PMT readout is foreseen to be deployed in 2026, while the one with SiPM readout at even lower temperature is under active R\&D.
An upgrade to 700\,kg in a modularized design is under study.
  
\item \textbf{Liquid Argon:}  The single-phase liquid argon detector, called CENNS-10, with a fiducial mass of 24\,kg, was located at the end of the Neutrino Alley hallway at a distance of 27.5\,m to the target \cite{Tayloe:2017edz}. An energy threshold of 20\,keV$_{nr}$ was achieved by reading out the scintillation light from the argon with two PMTs. This threshold was possible due to a light yield of $\sim$4.5\,photoelectrons per keV ionization energy. In 2021, CEvNS was observed with a 3.5\,$\sigma$ significance for an exposure of 6.1\,GWh \cite{COHERENT:2020iec}. Since then, three times more statistics has been collected following this first result and analysis is underway. Almost 500 CEvNS events are expected in the full dataset.  
To further increase the statistics, the setup will be replaced by COH-Ar-750, a larger liquid argon detector with a similar design. The light signals will be read out with 122 PMTs with a nominal quantum efficiency of 20\%. An expected fiducial mass of 476\,kg occupies a similar footprint in Neutrino Alley.
More than 5000 CEvNS events and more than 500 charged-current events are expected within one SNS year. 
A measurement of the $\nu_e$-Ar charged-current cross section will be achieved, which has never been measured experimentally in the tens-of-MeV range. The search for inelastic neutrino interactions at higher energies far above the region of interest for CEvNS will benefit from an upgraded shield design including a muon veto.

The new cryostat is currently under commissioning at ORNL. Beyond the scaling up of CENNS-10, there is also an additional proposal for a liquid argon time projection chamber (LArTPC) with a fiducial mass of 250\,kg and a pixelated readout, making use of similar technology as DUNE. This detector has potential to significantly reduce uncertainties on supernova-burst measurements with DUNE \cite{DUNE:2023rtr} by studying the final states of this reaction.

Furthermore, the CENNS-10 cryostat can be repurposed to be filled with liquid neon to include a further light isotope next to Na in the $N^{2}$ test of the standard-model cross section.

\item \textbf{Germanium:}
The COHERENT germanium setup, called Ge-Mini, consists of eight high-purity point contact inverse coaxial detectors with a mass of $\sim$2.2\,kg each. The liquid nitrogen-cooled detectors are located at an average distance of 19.2\,m to the neutrino source at the same location as the former CsI detector. With this setup and an exposure of 4.58\,GWh,  CEvNS was detected on germanium for the first time in 2023 with a significance of 3.9\,$\sigma$ \cite{COHERENT:2025vuz}. HPGe detectors have a very good intrinsic energy resolution and very low internal background, which combined with the short SNS beam pulses, results in an excellent signal to background ratio of about one to one and better.
While the previous detections were achieved with scintillation light, in this case CEvNS was measured with recoil-induced ionization energy. The threshold of 6.7\,keV$_{nr}$ and the achieved background level are the lowest within COHERENT out of the three CEvNS results. This first detection is fully statistics-dominated. More than twice as much data were collected in 2025.
For this dataset, with improved analysis techniques, the energy threshold was lowered even further to 2.5\,keV$_{nr}$ and the background in the region of interest was reduced by more than 50\% \cite{geminitemp}.   
All factors combined lowered the statistical uncertainty down to the order of 10\%. A total significance of $\sim$10\,$\sigma$ was achieved \cite{geminitemp}; the result is in 1\,$\sigma$ agreement with the standard model as shown in Fig. \ref{fig:cevnscrosssection}.

Beyond a continuous collection of more data, the shield has the capacity for four more detectors to be acquired in the future. In the further future, 200\,kg of HPGe detectors in a redesigned cryostat can enable sub-percent-level CEvNS measurements.

\item \textbf{NaI:}  The tonne-scale COHERENT NaI effort, NaIvETe, aims at a detection of CEvNS on sodium and a high statistics detection of charged-current interactions on iodine.
The first phase will consist of five modules and a total mass of 2.4\,t, followed by the addition of two more modules with a total mass of 3.5\,t. The single modules are constructed using 63 repurposed 7.7\,kg NaI[Tl] crystals with PMT light collection. Sodium is currently the lightest nucleus deployed in Neutrino Alley to look for CEvNS. A 3~$\sigma$ measurement per year of operation at the SNS is expected assuming a 13\,keV$_{nr}$ threshold.
COHERENT leverages here the experience from the NaIvE-185\,kg detector, which was used to study charged-current interactions on $^{127}$I in a higher energy range of 10-55\,MeV, above the region of interest for CEvNS. The interaction was detected with a significance of 5.8\,$\sigma$ \cite{COHERENT:2023ffx}, but the determined cross section is low by 41\% in comparison to the MARLEY prediction~\cite{gardiner2021simulating}.
Currently, four of the five modules are deployed in Neutrino Alley and detailed background studies are ongoing. The five-module setup is expected to be completed within this year. 

\item \textbf{Water-based Cherenkov Detectors} 
The main systematic uncertainty that all COHERENT detectors share is the uncertainty on the neutrino flux, amounting to 10\% as evaluated in Ref.~\cite{COHERENT:2021yvp}. To reduce this uncertainty, the D$_2$O Cherenkov detector measures the neutrino charged-current cross section on deuterons~\cite{COHERENT:2021xhx}. This cross section is theoretically understood at the few-percent level.
Module 1 (M1), a 549\,kg mass of D$_2$O in an acrylic vessel surrounded by 10\,cm of regular water as a ``tail catcher" in a stainless steel vessel, is read out with PMTs at a distance of 19.3\,m to the neutrino source. The detector is currently collecting data and is expected to reduce the flux uncertainty to 3-5\% within the next five years of operation.
Neutrino inelastic interactions on oxygen in the few tens-of-MeV range have never been measured and are relevant for interpretation of subdominant supernova-burst signals in Super-K and Hyper-K.  A second module (M2) is already built and under commissioning to be initially filled with light water; it will enable an inelastic neutrino-oxygen cross section measurements.  
The plan is to eventually fill it with heavy water.  Furthermore, a dedicated light water detector for measurement of $\nu_e$ interactions on oxygen is also under consideration.

\item \textbf{Other inelastic-interaction detectors:}  Several COHERENT subsystems investigate inelastic interactions on heavy nuclear targets: lead and iron ``neutrino cubes," lead glass and metal sandwich detectors, and a thorium-based detector for neutrino-induced fission measurements.  
  
Charged-current interactions of neutrinos on lead can produce ejected neutrons -- referred to as neutrino-induced neutrons, or NINs.
These are a background to CEvNS, as detectors tend to be surrounded by lead shields. In order to characterize NINs,
lead and iron ``neutrino cubes" consisting of Pb and Fe targets and liquid scintillator detectors were deployed in Neutrino Alley and took several years of data. The analysis of the Pb data rejected the null hypothesis of no neutrino-induced neutrons (NINs) by 1.8\,$\sigma$ \cite{COHERENT:2022fic}. The determined cross section corresponds to only 29\% of the value predicted by MARLEY. The measurement confirmed that the NIN background is subdominant for SNS CEvNS detection.     

A follow-on lead glass detector (NEP-ton) is in the process of being deployed in 2026 to evaluate the Pb NIN cross section to higher precision. Another planned detector is a ``metal sandwich," consisting of lead (and potentially other metal) layers interspersed with plastic scintillator plates. 

The NuThor detector aims at measuring neutrino-induced nuclear fission in thorium.  The setup consists of 52\,kg of thorium in a lead shield. The detectable signature of the neutrino-induced fission process is a large multiplicity of neutrons. These neutrons are moderated and captured in the surrounding Gd-doped water bricks and the resulting deexcitation gamma rays are registered with NaI detectors. A machine-learning based analysis approach extracts events with high neutron multiplicity. The unblinding of the first dataset showed a preference of 2.4\,$\sigma$ for the presence of neutrino-induced fission~\cite{Johnson:2024vsi}. More data are currently being collected by the NuThor detector.

\end{itemize}

\section{Physics Reach}
 
CEvNS and inelastic interaction detections at the SNS are a gateway to both beyond-the standard-model particle physics and to nuclear physics. Relevant physics signatures are modifications of the rate and recoil spectrum shape from the standard-model prediction, for which there is very well-understood flavor, energy and time structure from the neutrino source.

Physics that can be derived from the COHERENT data include:

\begin{itemize}
    \item {\bf Non-standard neutrino interactions (NSI)}: Beyond-the-standard model (BSM) physics can be introduced to CEvNS in an effective field theory approach. The non-standard coupling strengths of neutrinos to quarks are described by a matrix of flavor-dependent $\varepsilon$ parameters. Current COHERENT results already strongly limit the allowed parameter space (e.g., \cite{COHERENT:2017ipa,COHERENT:2020iec,Papoulias:2019xaw,Giunti:2019xpr,Liao:2024qoe}). Future detections will further improve the sensitivity and help to resolve the LMA and LMA-dark solar oscillation parameter ambiguity~\cite{COHERENT:2023sol}. The COHERENT NSI sensitivities are complementary to the LHC NSI sensitivies \cite{Lozano:2025ekx}.  
    \item {\bf Light mediators}: Another way to look for new physics is via light vector or scalar mediators, added to the standard model and CEvNS in a similar role as the Z boson. Current and future COHERENT detections set limits in the mass/coupling constant parameter space comparable to tonne-scale neutrino detectors (e.g. \cite{AtzoriCorona:2022moj}). 
    \item {\bf Weak mixing angle}: The weak mixing angle $\theta_W$ has been measured to a high precision at the TeV scale at colliders, but not in the tens-of-MeV momentum-transfer range. A precision detection of CEvNS enables such a measurement and any deviations from the standard-model prediction could point towards new physics~\cite{AtzoriCorona:2024vhj}. A precision of 2\% can be reached for 50\,kg of Ge with a conservative threshold of 5\,keV$_{nr}$.  
    \item {\bf Sterile neutrinos}: All systems of COHERENT combined can be considered a short-baseline experiment covering a distance of 19.2\,m to 28\,m to the SNS target. This enables COHERENT to test neutrino oscillations in the range of ${\Delta}m^2_{41}$ amounting to 0.4-3.4\,eV$^2$ using neutral-current disappearance~\cite{COHERENT:2022nrm}.   These investigations benefit from the small theoretical uncertainties on the CEvNS cross section. The electron-neutrino charged-current interaction detections will enhance the dataset for sterile neutrino studies.  In the further future, the STS will provide multiple baselines and allow cancellation of detector systematics by observation of spectrum modulation in a single detector.
    \item {\bf Accelerator-produced dark matter}: At the SNS beam energy, sub-keV dark matter searches can be carried out. It is assumed that hidden sector particles coupling to the dark matter are produced by proton interactions in the SNS target (beam dump). The resulting dark matter will cause recoils in the COHERENT detectors comparable to CEvNS, but with a different timing structure. In \cite{COHERENT:2021pvd} and \cite{COHERENT:2022pli} limits for hidden-sector particles kinetically mixing with standard-model quarks or photons are derived for the full CsI dataset. Future data will test the theoretically motivated relic abundance lines for both scalar and Majorana fermion dark matter over a wide parameter space. It is also possible to look for axion-like particles in the COHERENT data \cite{CCM:2021jmk}.    
    \item {\bf Nuclear physics}: The nuclear form factor in the CEvNS cross section describes the impact of the spatial distribution of the nucleons in the nucleus on the interaction rate and recoil spectrum. At the SNS neutrino energies, it is possible to use the CEvNS detections to derive information on the form factor and ultimately average neutron radius $R_{n}$ (e.g., \cite{AtzoriCorona:2024vhj,Coloma:2020nhf}). The parameter $R_{n}$ as well as the ``neutron skin" are relevant for astrophysics in understanding gravitational-wave merger events and the equation of state of neutron stars. COH-Ar-750 and CryoCsI will measure $R_{n}$ with a precision of 5\% or better within three years of operation. For the form-factor determination, measurements of the same isotope at SNS and at reactor sites are highly desirable due to the different momentum transfer regime. The reactor experiments will set the normalization, while the impact of the form factor can be observed in the COHERENT data (as visible in the difference between the green and black line in Fig. \ref{fig:cevnscrosssection}). With the very recent first CEvNS detection at a reactor site \cite{Ackermann:2025obx}, suitable datasets will be achieved in the relatively near future that will help to disentangle nuclear effects from potential beyond-the-standard-model physics.  
    \item {\bf Supernovae}: Neutrinos emitted in core-collapse supernova explosions have energies of a few tens-of-MeV and thus will result in both CEvNS and inelastic interactions in the COHERENT detectors. Because CEvNS is flavor-blind to a good approximation, it enables a complete estimate of all emitted neutrinos, while multi-kilotonne-scale experiments such as DUNE, Hyper-K, and JUNO will be primarily sensitive to only the $\nu_e$ and $\bar{\nu}_e$ flavor components of the supernova flux.  Cross-section measurements at SNS will vastly improve the interpretation of supernova neutrinos in large-scale experiments using the well-understood ``artificial supernova" stopped-pion source at the SNS. 
    Additionally, COHERENT’s tonne-scale detectors can also directly observe nearby supernovae with O(10) events expected from a core-collapse supernovae at 10\,kpc.  
\end{itemize}

\begin{figure}[htbp]\centering
  \includegraphics[width = 0.48\textwidth]{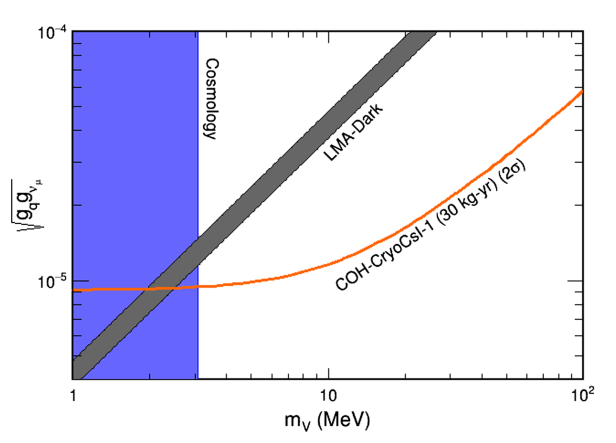 }
  \includegraphics[width = 0.48\textwidth]{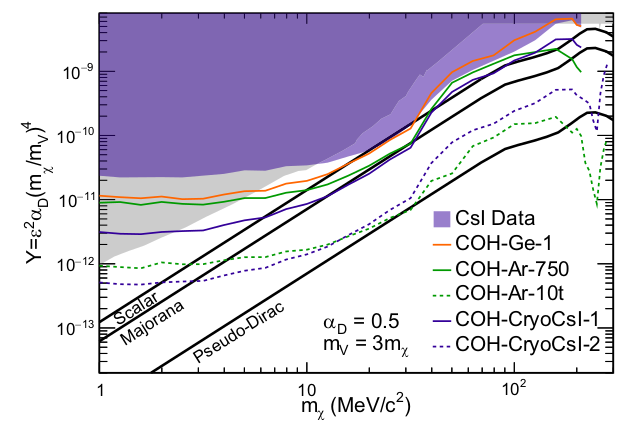 }
  \caption{Left: using CEvNS to probe the LMA dark solution to the solar neutrino problem with CryoCsI with 10\,kg mass and 30\,kg-yr exposure. Right: projected sensitivities for exclusion limits on sub-GeV dark matter.  Figures from Ref.~\cite{COHERENT:2023sol}.}
  \label{fig:bsmphysics}
\end{figure}

The COHERENT collaboration has released the data from the CsI \cite{Akimov:2018vzs, akimov2022measurementdatarelease} and LAr \cite{COHERENT:2020ybo} CEvNS results. A release of the Ge dataset is in progress. Data releases not only assure reproducibility, but also enable the theory and phenomenology community to test their models and carry out detailed investigations of BSM physics.

\section{Conclusion}

The CEvNS interaction enables precision tests of the standard model due to the large cross section for a neutrino interaction and the small nuclear uncertainties. A high-statistics CEvNS detection  does not require kilotonne scale detectors. The SNS is an ideal place to detect CEvNS due to the high beam power and the sharply pulsed beam that enables a drastically reduced steady-state background.

The COHERENT experiment has already successfully detected CEvNS with tens-of-kg detector mass on CsI, Ar and Ge nuclei. In the next years, many more CEvNS events will be collected by lowering the energy thresholds and increasing the mass to 750\,kg for the single-phase liquid argon detector. Na and Ne are the next new CEvNS targets planned. The major systematic uncertainty of the neutrino flux of 10\% will be addressed by the D$_2$O detector, which will ultimately enable a few-percent-level evaluation of the cross section within the next $\sim$5\,years. 
With these data, detailed studies of beyond-the-standard-model physics can be carried out including limits on NSIs, light mediators and accelerator-generated dark matter. Nuclear properties such as the neutron form factor can be evaluated directly, also leveraging here complementary information that will be provided by reactor experiments at lower energies. Precise knowledge of the CEvNS cross section will also ensure that WIMP dark matter experiments will understand their neutrino background as they increase their sensitivity and move into the ``neutrino fog" region.

Furthermore, COHERENT measures inelastic interactions in the tens-of-MeV range. The precision of COHERENT's existing measurements on I and Pb will continue to improve, and more detections will follow on Ar and O. These results will ensure that large-scale neutrino observatories can successfully interpret their low energy astrophysical data.

\section{Acknowledgements}

COHERENT is funded by a variety of sources. The COHERENT Collaboration acknowledges the generous resources provided by the ORNL Spallation Neutron Source, a DOE Office of Science User Facility. Laboratory Directed Research and Development funds from ORNL also supported this project. We acknowledge support from U.S. Department of Energy Office of Science and the National Science Foundation.  
We also acknowledge support from the Alfred P. Sloan Foundation, the Consortium for Nonproliferation Enabling Capabilities, and the Korea National Research Foundation (No. NRF 2022R1A3B1078756). This research used the Oak Ridge Leadership Computing Facility, which is a DOE Office of Science User Facility. We also acknowledge support from Ministry of Science and Higher Education of the Russian Federation, Project ``Studying physical phenomena in the micro- and macro-world to develop future technologies," FSWU-2026-0010.

\bibliographystyle{JHEP}
\bibliography{ref}
\end{document}